# EPR Imaging of Metallic Lithium

Arvid Niemöller [1], Peter Jakes [1], Rüdiger-A. Eichel [1,2], and Josef Granwehr [1,3]

[1] Forschungszentrum Jülich GmbH, Institut für Energie und Klimaforschung (IEK-9), 52425 Jülich, Germany
[2] Institut für Physikalische Chemie, RWTH Aachen University, 52056 Aachen, Germany
[3] Institut für Technische und Makromolekulare Chemie (ITMC), RWTH Aachen University, 52056 Aachen, Germany

**Abstract**

We present a sensitive characterization method to image the microstructure of lithium deposits in lithium-ion battery components by Conduction Electron Paramagnetic Resonance Imaging (CEPRI). The versatility of the method is demonstrated for both, imaging surface-patterns of thick lithium metal anodes, as well as obtaining high-resolution images of lithium dendrites formed inside a separator with several micrometre pixel size. The determined spatial distribution of dendrites may then serve as an indicator of the current density distribution and, therefore, yields most valuable information for battery cell design. Accordingly, this method shows its capabilities in a research area where electron paramagnetic resonance imaging (EPRI) has not been utilized so far.

## Introduction

Lithium metal anodes are considered as an ideal anode material for rechargeable lithium-ion batteries owing to the comparatively high theoretical specific capacity of 3860 mA h g$^{-1}$ and the lowest negative electrochemical potential of −3.040 V *vs.* the SHE. Despite these attractive properties, a practical application in lithium-ion cells is hampered by an hitherto non-controllable formation of lithium dendrites[1] and a limited Coulombic efficiency during Li deposition/stripping.[2] To guarantee safe and efficient operation of lithium metal anodes, a fundamental knowledge about the underlying mechanisms that promote dendritic lithium growth when depositing lithium is mandatory. Of particular interest is to quantitatively investigate the impact of cell architecture and geometry, which imposes significant variations in current-density distribution, on the dendrite formation characteristics. This is of particular interest for the design of battery cells for fast charging purposes.

The cell geometry generally imposes an intrinsic inhomogeneity in current density. Accordingly, lithium deposition is inhomogeneous, and therefore dendrites grow at preferred sites.[3] I.e. dendrites are formed at sites of maxima of current density.[4] Therefore imaging of dendrites in separators can give evidence for the current distribution in a lithium ion battery,[5] whose visualization is a major battery design criterium, including battery-pack shape, cell geometry and active material arrangement.[6,7]

Standardly used imaging techniques are FIB-SEM, MRT or CT. Furthermore, optical analyses like Raman, IR or fluorescence imaging are advanced methods for image recording. In terms of metallic lithium structures in the micrometre range, these methods suffer from lateral resolution, maximum sample size, a destructive analysis or shielding. Furthermore, quantitative analyses are then hardly to obtain.

Spin density mapping with electron paramagnetic resonance imaging (EPRI) is an advanced technique[8] that is mostly used for non-conductive, paramagnetic samples, often in combination with spin probes. Examples include chemical, physical and biological problems.[9] So far, EPRI of metallic lithium has not been reported, even though conduction EPR (CEPR) of this material is well established.[10,11] The most recent works concerning lithium EPR concentrated on lithium doped ceramics,[12] the identification of mossy lithium depending on the state of charge of a lithium-ion battery,[1] and the quantitative analysis of lithium plating.[13]

EPR investigations of conductive samples are generally complicated by the skin effect. If the sample thickness $d$ is much larger than the skin depth of the microwave frequency, the first derivative EPR spectrum shows a Dysonian lineshape that can be approximated as phase shifted Lorentzian.[10,14] It has been demonstrated that this effect can influence conduction EPR imaging (CEPRI) and additional measurements are necessary for complete data sets.[15,16]

The pixel size $\Delta z$ of EPR images correlates with the EPR peak-to-peak linewidth $\Delta H_{pp}$, i.e. narrow lines allow higher resolution images. In addition, high gradients $G$ are also preferable for high resolution:

$$\Delta z \propto \frac{\Delta H_{pp}}{G} \tag{1}$$

For this reason, only specific samples are suitable for EPRI and CEPRI to generate a useful image. For metallic lithium, the linewidth strongly depends on the morphology. Thick lithium, with a skin depth $\delta \ll d$, shows a typical Dysonian lineshape with a peak-to-peak linewidth of ca. 0.15 mT. Mossy lithium that is formed during electrochemical cycling has a linewidth of ca. 0.03 mT and, finally, dendritic lithium in, e.g., battery separators shows a linewidth of ca. 0.005 mT (Fig. 1). Dendritic lithium is of similar dimensions as the EPR microwave skin depth, thus a Lorentzian line shape is obtained. Therefore, dendritic lithium is an ideal candidate for CEPRI as it combines a high resolution due to narrow linewidth and a Lorentzian line shape that simplifies data recording and analysis, also in a quantitative manner.

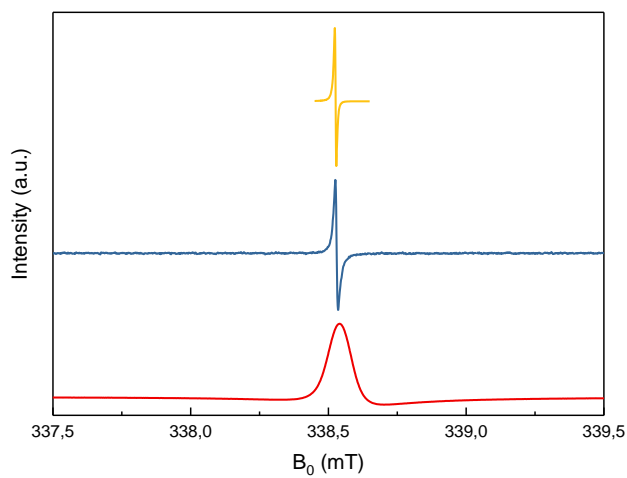

**Figure 1. First derivative EPR signal for lithium with different morphologies. The peak-to-peak linewidth is minimal for dendritic lithium with ca. 0.005 mT (yellow) showing a Lorentzian line shape, increases for mossy lithium to 0.03 mT (blue) and is maximal for bulk lithium with ca. 0.15 mT (red) showing a Dysonian lineshape.**

## Results & Discussion

To demonstrate CEPRI of metallic lithium, different 380 µm thick, flat lithium pieces were prepared. Samples were perforated on one side with a pyramidal structure and compared with an unperforated sample. Furthermore, a quadratic sample with a sharp edge was analysed as well (Fig. 2, bottom row).

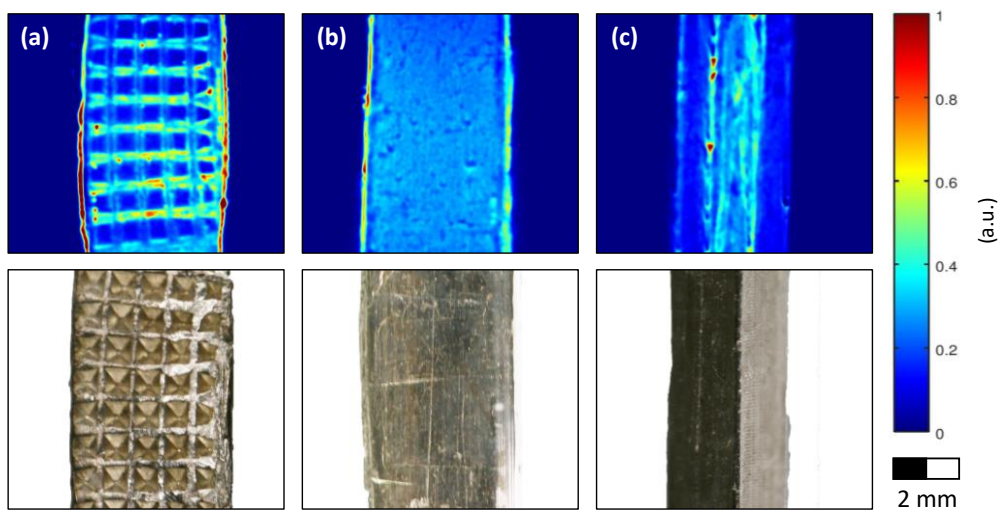

**Figure 2.** Thick metallic lithium pieces perforated mechanically with different stamps. Top row: EPR images. Bottom row: photographs of the samples. (a) Sample with single-sided pyramidal pattern and a flat surface on the other side. (b) Two flat surfaces. (c) Lithium piece with a square cross section and a slight inhomogeneous surface.

After spectra acquisition, image reconstruction was done by reference deconvolution and filtered back projection.[17] Since microwave phase was found to be constant at the lithium surface, reference deconvolution did not show any phase dependent artefacts. In more detail, the reference deconvolution of projections is used to obtain the absorption spectrum from the standardly recorded first derivative spectrum in EPR experiments.[18] Subsequently, a filtered back-projection, i.e. an inverse Radon transform, provides the spatial-spatial image.[19] A huge advantage of the reference deconvolution is an increasing signal to noise ratio and thereby an increase of resolution in contrast to a simple integration method.[20] In case of a multicomponent system with EPR signals that show different phase shifts, a deconvolution would result in a disturbed unfolded projection and different approaches for image reconstruction are necessary.

The shape of the pyramids is shown in Fig. 2a, where one side of the lithium was perforated. The quadratic valleys are clearly visible and the bars in between show a higher intensity, also being more exposed. Thus, the bars generate an additional shielding as the valleys show a weak EPR signal. The flat sample in Fig. 2b shows a constant intensity except for the edges. In comparison to the optical image, EPR intensity is amplified at the edges even though the surface appears to be homogenous. The edge for the square sample is of low visibility, only surface inhomogeneities have a higher EPR intensity (Fig. 2c).

Lithium intensity in all images of Fig. 2 correlates with the magnetic field induced at a certain point and therefore, with the microwave field distribution. Since microwave absorption is different for fine spikes, edges and flat planes, even though the skin depth is below the sample thickness, the EPR intensity gives information about the induced magnetic field, but a quantification of signal intensities due to shielding is rather difficult. Especially for edges, the induced magnetic field is higher as it is visible in Fig. 2b. Furthermore, the sensitivity towards geometric patterns allows resolving macroscopic structures, here in the sub-millimetre range. In terms of acquisition parameters, resolution of images with a zero-gradient linewidth of 0.15 mT and an applied gradient of 5 mT cm$^{-1}$ is in the range of 50 μm.

Since the image pixel size is proportional to the EPR linewidth, a much higher resolution was obtained for imaging of lithium dendrites. Therefore, dendrites were grown in a battery separator and the EPR spectrum shows a dendritic lithium signal with a narrow, Lorentzian lineshape of 0.005 mT width (Fig. 1).

The separator was positioned parallel to the image plane inside the resonator and the recorded and reconstructed image is shown in Fig. 3 resulting in a resolution of several micrometres. The Lorentzian lineshape indicates that the dendrite dimensions are below the microwave skin depth. Therefore, shielding effects are not dominant and image intensities can be interpreted quantitatively in terms of local spin density (or concentration). Moreover, concerns for image reconstruction, i.e. the reference deconvolution, are not relevant even though a conductive material is investigated. The microwave phase is not influenced by the skin effect and further, is constant throughout the sample. Since the image was recorded in two dimensions and the separator had an original thickness of 150 µm, the dendrite signal is an accumulation through the height of the separator for each spot in the imaging plane. Therefore, a high intensity suggests high amounts of lithium dendrites, i.e. long dendrites, while low intensities are rather short dendrites.

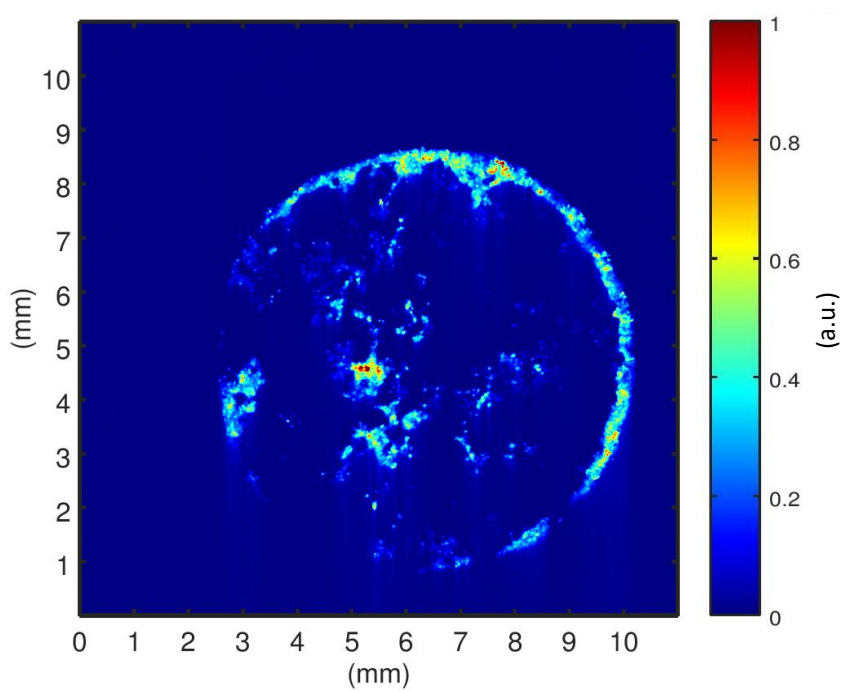

**Figure 3. EPR image of lithium dendrites grown in an 8 mm glass fibre separator. The appearance of dendrites is preferably found at the edge of the separator. In addition, an intense dendrite signal is found in its centre.**

Fig. 3 exhibits that dendrite growth is not homogenously distributed across the separator. Certain spots are more intense, implying a higher concentration or length of dendritic lithium species. The dendrite distribution provides a hint regarding the current distribution inside the battery. In detail, the result implies that this distribution is not homogeneous in between the electrodes. As shown in literature, the electric field in such an electrode arrangement is higher at the edges, which is supported by the observed lithium dendrite growth.[5] In addition, the most intense dendrite spot is found in the centre of the separator. Most probably, this dendrite structure caused the observed short circuit after 200 battery cycles. As large areas are not covered with dendrites, it shows that the dendrite distribution is not randomly. Further investigations of cycle dependent dendrite distribution will help to understand the patterns of dendrite growth.

**Conclusion**

In conclusion, we have shown the capabilities of CEPRI to investigate metallic lithium samples. Thick lithium samples with textured surface and a battery separator containing dendritic lithium generated by electrochemical cycling were analysed. From these images, structural surface analyses and the determination of distributions for conductive lithium species are possible, offering new insights for e.g. battery application. In a next step, the quantification of dendritic lithium by CEPRI will be developed. Furthermore, the signal composition for multicomponent systems will be targeted.

## Methods
**EPR Spectroscopy**

CEPRI experiments were performed on a Bruker E540 Elexsys X-band spectrometer and a 4108 TMHS resonator. Imaging of thick lithium samples was done with a 5 mT cm$^{-1}$ gradient. The modulation amplitude was set to 0.03 mT at a frequency of 100 kHz for thick lithium samples with a microwave power of 0.63 mW. Sweep width was 8 mT with 20 s sweep time. Dendrite imaging was done with a 2.5 mT cm$^{-1}$ gradient and a modulation frequency of 10 kHz at 0.001 mT modulation amplitude. A single projection was obtained in 240 s with a sweep width of 4 mT. Microwave power was set to 1 µW and 402 projections were recorded in both experiments.

**Image reconstruction**

Image reconstruction was done by a reference deconvolution of the projections applying a Gaussian window in GNU Octave (version 4.2.1). For this method, projections and a reference spectrum with zero gradient were recorded. Afterwards, a filtered back projection with a cosine filter was used.

**Sample preparation**

Structured lithium metallic samples were prepared with stamps manufactured in-house. Lithium metal was purchased from Sigma Aldrich with an assigned purity of 99.9 %.

A separator containing dendritic lithium was formed in a Swagelok battery cell. In more detail, a linear battery electrode arrangement, where two circular stamps were positioned parallel and the active materials were placed in between, was used applying currents of 10 mA, for 200 cycles until a short circuit was detected. Thereby, metallic lithium was cycled against metallic lithium utilizing a glass fibre separator from VWR and electrolyte from Sigma Aldrich (LP30). Afterwards, the separator was extracted and fixed in a sample holder under argon atmosphere for EPR imaging. Mossy lithium was harvested from the metallic lithium electrodes used for the formation of dendritic lithium.

All lithium samples were handled under argon atmosphere and a gas tight container was utilized for EPR measurements.

## Acknowledgement
We thank the mechanical workshop of the Department of Electronic Systems (ZEA-2, Forschungszentrum Jülich) for fabrication of lithium stamps.